\documentstyle[preprint,aps]{revtex}
\input epsf.tex
\def\DESepsf(#1 width #2){\epsfxsize=#2 \epsfbox{#1}}
\begin{document}
\preprint{\vbox{\hbox{OITS-576}}}
\draft
\title{A Method for Determining CP Violating Phase $\gamma$\footnote{Work
supported in part by the Department of Energy Grant No.
DE-FG06-85ER40224.}}
\author{N.G. Deshpande, and Xiao-Gang He}
\address{Institute of Theoretical Science\\
University of Oregon\\
Eugene, OR 97403-5203, USA}

\date{May 1995}
\maketitle
\begin{abstract}
A new way of determining the phases of weak amplitudes in charged $B$ decays
based on SU(3) symmetry is proposed. The CP violating phase $\gamma$ can now be
determined without the previous difficulty associated with electroweak
penguins.
\end{abstract}
Detection of CP violation and verification of the unitarity triangle of the CKM
matrix is a major goal of $B$ factories\cite{1}. Decisive information about the
origin
of CP violation will be obtained if the three phases $\alpha =
\mbox{arg}(-V_{td}V_{tb}^*/V_{ub}^*V_{ud})$, $\beta =
\mbox{arg}(-V_{cd}V_{cb}^*/V_{tb}^*V_{td})$ and $\gamma =
\mbox{arg}(-V_{ud}V_{ub}^*/V_{cb}^*V_{cd})$ can be independently measured
experimentally\cite{2}. The sum of these three phases must be equal to $180^o$
if
the Standard Model with  three generations is the model for CP violation.
There have been many studies to measure these phases. The phase $\beta$ can be
determined unambiguously by measuring time variation asymmetry in $\bar B^0
(B^0) \rightarrow \psi K_S$ decay rates\cite{2}. The phase $\alpha$ can be
measured in $B^-\rightarrow \pi\pi$ and
$B\rightarrow \rho \pi$ decays\cite{3,4}. In these decays, there are
contributions from
the tree and the penguin (both strong and electroweak) amplitudes. The methods
proposed in Refs.\cite{3,4} are
valid even if the strong penguin contributions are included. The inclusion of
the electroweak penguin contributions may contaminate the result. However,
because the electroweak
penguin effects are small in this case, the error in $\alpha$ determination
are small. Several methods using $\Delta S = 1$ $B$ decays to measure
the phase $\gamma$ had been proposed\cite{5,6}.  Most had
assumed that the effects from electroweak penguin
could be neglected. It has been recently shown by us\cite{7} that
this assumption
is badly violated for top quark mass of order 170 GeV.
For $\Delta S = 0$ hadronic $B$ decays, the strong penguin effects are
much smaller than the leading tree contributions, and thus electroweak penguin
effects which are even smaller can be safely ignored. In $\Delta S = 1$ decays,
because of the large enhancement factor $|V_{tb}V_{ts}^*/V_{ub}V_{us}^*|\approx
55$, the strong penguins dominate and the electroweak penguin effects are
comparable to the tree contributions. This invalidates methods proposed in
Refs.\cite{5,6}. In this letter we give further consideration to measuring
$\gamma$ using $\Delta S = 1$ decays, although other methods have been
discussed in the literature\cite{2,gamma}.

An interesting method has recently been
proposed by Hernadez, Gronau, London and Rosner\cite{8} using SU(3) relations
between the decay amplitudes for $B^-\rightarrow \pi^0 K^-$, $\pi^- \bar K^0$,
$\pi^0\bar K^0$, $\pi^+ K^-$, $B^-\rightarrow \pi^-\pi^0$, and
$B_s \rightarrow \eta \pi^0$. This method requires the reconstruction of the
quadrangle from $B^-\rightarrow \pi^0 K^-$, $\pi^- \bar K^0$, $\pi^0\bar K^0$,
$\pi^+ K^-$ decays. In order to do so, one not only needs to measure all the
four $B\rightarrow \pi K$ decay amplitude but also needs to measure the rare
decay
amplitude $B_s \rightarrow \eta \pi^0$. It has been shown that this last decay
is a pure $\Delta I = 1$ transition, with the dominant contribution from the
electroweak penguin. However, the branching ratio is extremely small
$O(10^{-7})$\cite{9}. In this letter we propose a new method to measure
$\gamma$ using  $\Delta S = 1$ decays $B^-\rightarrow
\pi^- \bar K^0$, $ \pi^0 K^- $, $\eta K^-$, and the $\Delta S = 0$ decay
$B^-\rightarrow \pi^-\pi^0$, which relies on SU(3) symmetry. This method
is free from the electroweak penguin contamination problem, and further,
 all decays involved have relatively large  ($O(10^{-5}$)) branching ratios.
More importantly these measurements can in principle be carried out at present
facilities like
CLEO or CDF/D0.

In the SM the most general effective Hamiltonian for
hadronic $B$ decyas can be written as follows:
\begin{eqnarray}
H_{eff}^q &=& {G_F\over \sqrt{2}}[V_{ub}V^*_{uq}(c_1O_1^q + c_2 O_2^q) -
\sum_{i=3}^{10}V_{tb}V^*_{tq} c_i
O_i^q] +H.C.\;,
\end{eqnarray}
where the $O_i^q$ are
defined as
\begin{eqnarray}
O_1^q &=& \bar q_\alpha \gamma_\mu(1-\gamma_5)u_\beta\bar
u_\beta\gamma^\mu(1-\gamma_5)b_\alpha\;,\;O_2^q =\bar q
\gamma_\mu(1-\gamma_5)u\bar
u\gamma^\mu(1-\gamma_5)b\;,\nonumber\\
O_{3,5}^q &=&\bar q \gamma_\mu(1-\gamma_5)b
\bar q' \gamma_\mu(1\mp\gamma_5) q'\;,\;\;\;\;\;\;\;O_{4,6}^q = \bar q_\alpha
\gamma_\mu(1-\gamma_5)b_\beta
\bar q'_\beta \gamma_\mu(1\mp\gamma_5) q'_\alpha\;,\\
O_{7,9}^q &=& {3\over 2}\bar q \gamma_\mu(1-\gamma_5)b  e_{q'}\bar q'
\gamma^\mu(1\pm\gamma_5)q'\;,\;O_{8,10}^q = {3\over 2}\bar q_\alpha
\gamma_\mu(1-\gamma_5)b_\beta
e_{q'}\bar q'_\beta \gamma_\mu(1\pm\gamma_5) q'_\alpha\;,\nonumber
\end{eqnarray}
Here $q'$ is summed over u, d, and s. For $\Delta S = 0$ processes,
$q = d$, and for $\Delta S = 1$ processes, $q =s$. $O_{2}$, $O_1$ are the tree
level and QCD corrected operators. $O_{3-6}$ are the strong gluon induced
penguin operators, and operators $O_{7-10}$ are due to $\gamma$ and Z exchange,
and the ``box'' diagrams at one loop level (electroweak penguin). The Wilson
coefficients $c_i$ are defined
at the scale of $\mu \approx
m_b$ which have been evaluated to the next-to-leading order in QCD\cite{10}.
In the above we have neglected small contributions from u and c quark loop
contributions proportional to $V_{ub}V_{uq}^*$.

We can parametrise the decay amplitude of $B$ as
\begin{eqnarray}
A = <final\;state|H_{eff}^q|B> = V_{ub}V^*_{uq} T(q) + V_{tb}V^*_{tq}P(q)\;,
\end{eqnarray}
where $T(q)$ contains the $tree$ contribution, while $P(q)$ contains $penguin$
contributions. SU(3) symmetry will lead to specific relations among $B$ decay
amplitudes.

SU(3) relations for $B$ decays have been studied by several
authors\cite{11,12}.
We will follow the notation used in Ref.\cite{12}.
The operators $Q_{1,2}$, $O_{3-6}$, and $O_{7-10}$ transform under SU(3)
symmetry as $\bar 3_a + \bar 3_b +6 + \overline {15}$,
$\bar 3$, and $\bar 3_a + \bar 3_b +6 + \overline {15}$, respectively. In
general,
we can
write the SU(3) invariant amplitude for $B$ to two octet pseudoscalar mesons
in the following form
\begin{eqnarray}
T&=& A_{(\bar 3)}^TB_i H(\bar 3)^i (M_l^k M_k^l) + C^T_{(\bar 3)}
B_i M^i_kM^k_jH(\bar 3)^j \nonumber\\
&+& A^T_{(6)}B_i H(6)^{ij}_k M^l_jM^k_l + C^T_{(6)}B_iM^i_jH(6
)^{jk}_lM^l_k\nonumber\\
&+&A^T_{(\overline {15})}B_i H(\overline {15})^{ij}_k M^l_jM^k_l +
C^T_{(\overline
{15})}B_iM^i_j
H(\overline {15} )^{jk}_lM^l_k\;,
\end{eqnarray}
where $B_i = (B^-, \bar B^0, \bar B^0_s)$ is a SU(3) triplet, $M_{i}^j$ is the
SU(3)
pseudoscalar octet, and the
matrices H represent the transformation properties of the operators $O_{1
-10}$.
$H(6)$ is a traceless tensor that is antisymmetric on its upper indices, and
$H(\overline {15} )$ is also a traceless tensor but is symmetric on its
upper indices. For $q=d$, the non-zero entries of the H matrices are given by
\begin{eqnarray}
H(\bar 3)^2 &=& 1\;,\;\;
H(6)^{12}_1 = H(6)^{23}_3 = 1\;,\;\;H(6)^{21}_1 = H(6)^{32}_3 =
-1\;,\nonumber\\
H(\overline {15} )^{12}_1 &=& H(\overline {15} )^{21}_1 = 3\;,\; H(\overline
{15} )^{22}_2 =
-2\;,\;
H(\overline {15} )^{32}_3 = H(\overline {15} )^{23}_3 = -1\;.
\end{eqnarray}
For $q = s$, the non-zero entries are
\begin{eqnarray}
H(\bar 3)^3 &=& 1\;,\;\;
H(6)^{13}_1 = H(6)^{32}_2 = 1\;,\;\;H(6)^{31}_1 = H(6)^{23}_2 =
-1\;,\nonumber\\
H(\overline {15} )^{13}_1 &=& H(\overline {15} ) ^{31}_1 = 3\;,\; H(\overline
{15} )^{33}_3 =
-2\;,\;
H(\overline {15} )^{32}_2 = H(\overline {15} )^{23}_2 = -1\;.
\end{eqnarray}
In terms of the SU(3) invariant amplitudes, the decay amplitudes $T(\pi\pi)$,
$T(\pi K)$ for
$\bar B^0 \rightarrow \pi \pi$, $\bar B^0 \rightarrow \pi K$ are given by
\begin{eqnarray}
T(\pi^-\bar K^0) &=& C^T_{(\bar 3)}
+A^T_{(6)} - C^T_{(6)} + 3A^T_{(\overline {15} )} -  C^T_{(\overline {15}
)}\;,\nonumber\\
T(\pi^0K^-) &=& {1\over \sqrt{2}} (C^T_{(\bar 3)}
+A^T_{(6)} - C^T_{(6)} + 3A^T_{(\overline {15} )} +7 C^T_{(\overline {15}
)})\;,\nonumber\\
T(\eta_8K^-) &=& {1\over\sqrt{6}}(-C^T_{(\bar 3)}
-A^T_{(6)} + C^T_{(6)} - 3A^T_{(\overline {15} )} +9 C^T_{(\overline {15}
)})\;,\nonumber\\
T(\pi^0\pi^-) &=& {8\over \sqrt{2}}C^T_{(\overline {15} )}\;,\nonumber\\
\end{eqnarray}
We also have similar relations for the amplitude $P(q)$. The corresponding
SU(3) invariant amplitudes will be denoted
by $A^P_i$ and $C^P_i$.
It is easy to obtain the following triangle relation from above:
\begin{eqnarray}
\sqrt{2} A(\pi^0 K^-) - 2A(\pi^- \bar K^0) = \sqrt{6} A(\eta_8 K^-)\;.
\end{eqnarray}

For the moment if we ingnore $\eta-\eta'$ mixing, it is clear that we can
construct this triangle from the experimentally measured rates for the
various $B^-$ decays. A similar triangle can also be constructed for the modes
of $B^+$ decay:
\begin{eqnarray}
\sqrt{2} \bar A(\pi^0 K^+) - 2\bar A(\pi^+ K^0) = \sqrt{6} \bar A (\eta_8
K^+)\;.
\end{eqnarray}

We shall now use a hypothesis that the tree contribution to the mode
$B^-\rightarrow \pi^- \bar K^0$ is negligible\cite{wolf}. This
 is varified in factorization approximation and had been assumed by
Ref.\cite{6,8}.
Further, if we work in Wolfenstein parametrization of the CKM matrix, the
amplitude $A(\pi^-\bar K^0)$
contains no weak phase. Then
\begin{eqnarray}
A(\pi^-\bar K^0) = \bar A(\pi^+K^0)\;.
\end{eqnarray}

We can now obtain the magnitude and relative phases of $A(\pi^0 (\eta_8) K^-)$
and $\bar A(\pi^0 (\eta_8) K^+)$ subject to two fold ambiguities
related to whether
the triangles for the $B^-$ and $B^+$ decays are on the same side (solution a)
or opposite side (solution b) of $A(\pi^- \bar K^0)$ as shown in Figure 1.

Now we construct two complex quantities (shown in Figure 1)
\begin{eqnarray}
B = \sqrt{2} A(\pi^0 K^-) - A(\pi^-\bar K^0) = 8(|V_{ub}V_{us}^*|e^{-i\gamma}
C^T_{\overline 15}
+|V_{tb}V_{ts}^*| C^P_{\overline 15})\;.
\end{eqnarray}
and
\begin{eqnarray}
\bar B = \sqrt{2} \bar A(\pi^0 K^+) -\bar A(\pi^+ K^0) =
8(|V_{ub}V_{us}^*|e^{i\gamma} C^T_{\overline 15}
+|V_{tb}V_{ts}^*| C^P_{\overline 15})\;.
\end{eqnarray}
Then
\begin{eqnarray}
B - \bar B = -i16|V_{ub}V_{us}^*|C^T_{\overline 15} \mbox{sin}\gamma\;.
\end{eqnarray}

To determine $\mbox{sin}\gamma$, we need a way of measuring $C^T_{\overline
{15}}$.  We achieve this by relating $C^T_{\overline 15}$ to
the amplitude $A(\pi^-\pi^0)$ for $B^-\rightarrow \pi^-\pi^0$. In the SU(3)
limit this decay amplitude is given by
\begin{eqnarray}
A(\pi^-\pi^0) = {8\over \sqrt{2}}(|V_{ub}V_{ud}^*|e^{-i\gamma} C^T_{\overline
15}
+|V_{tb}V_{td}^*| C^P_{\overline 15})\;.
\end{eqnarray}
Here the penguin contribution $C^P_{\overline {15}}$ arises from the
electroweak penguin only, and contributes
 less than 4\%\cite{7}.
We therefore obtain
\begin{eqnarray}
B - \bar B = -i2\sqrt{2} e^{i\delta^T_{\overline 15}}{|V_{us}|\over |V_{ud}|}
|A(\pi^-\pi^0)| \mbox{sin}\gamma\;,
\end{eqnarray}
where $\delta^T_{\overline 15}$ is the strong final state rescattering phase of
$C^T_{\overline 15}$. Thus the magnitude of $B-\bar B$ can be used to determine
$\mbox{sin}\gamma$. The phase of $B - \bar B$ gives us information of strong
phase $\delta^T_{\overline 15}$ relative to the strong phase of $A(\pi^- \bar
K^0)$. The two solutions for $B-\bar B$ in Figure 1
corresponding to a smaller value
(solution $(B-\bar B)_a$) and a larger value (solution $(B-\bar B)_b$) for
$\mbox{sin}\gamma$. The
larger value is likely to be ruled out because the resulting $\mbox{sin}\gamma$
may very well exceed unity.

Similarly, one can use the combination
\begin{eqnarray}
\tilde B &=& \sqrt{2} A(\pi^0 K^-) + \sqrt{6} A(\eta_8 K^-)
= 16(|V_{ub}V_{us}^*|e^{-i\gamma} C^T_{\overline 15}
+|V_{tb}V_{ts}^*| C^P_{\overline 15})\;;\nonumber\\
\tilde {\bar B} &=& \sqrt{2}\bar  A(\pi^0 K^+) +
\sqrt{6}\bar A(\eta_8 K^+)
= 16(|V_{ub}V_{us}^*|e^{i\gamma} C^T_{\overline 15}
+|V_{tb}V_{ts}^*| C^P_{\overline 15})\;,
\end{eqnarray}
to determine $\gamma$. In this case we have
\begin{eqnarray}
\tilde B - \tilde {\bar B} = -i4\sqrt{2} e^{i\delta^T_{\overline
15}}{|V_{us}|\over |V_{ud}|}
|A(\pi^-\pi^0)| \mbox{sin}\gamma\;,
\end{eqnarray}

The results obtained hold in the exact SU(3) limit.
SU(3) breaking effects in several ways will affect the above relations. These
include $\eta-\eta'$ mixing effect, the breaking effects in form factors and
mass differences.
Due to the $\eta-\eta'$ mixing effect to determine
$A(\eta_8 K^-)$, we need to determine the decay amplitudes $A(\eta K^-)$ and
$A(\eta' K^-)$. These amplitudes can be obtained from experiments. We can then
construct
\begin{eqnarray}
A(\eta_8 K^-) = \mbox{cos}\theta A(\eta K^-) + \mbox{sin}\theta A(\eta' K^-)\;,
\end{eqnarray}
where $\theta \approx 20^0$\cite{13} is the $\eta-\eta'$ mixing angle.
In principle we need to know the relative phase of $A(\eta'K^-)$ and
$A(\eta K^-)$. Since sin$\theta$ is small, this phase is important only if
$A(\eta' K^-)$ is much larger than $A(\eta K^-)$.
There are no reliable methods to evaluate the form factors at present. A
factorization approximation calculation indicates that
the large part of the  effect is due to
different decay constants and can be corrected by changing $A(\pi^-\pi^0)$ to
$(f_K/f_\pi)A(\pi^0\pi^-)$ in eqs.(15,16,17), and $A(\eta_8 K^-)$ to
$(f_K/f_\eta) A(\eta_8 K^-)$ in eqs. (8,9,16)\cite{14}.

A similar analysis can be carried out for $B^- \rightarrow \rho^0 K^{*-}\;,
\rho^-\bar K^{*0}\;, \omega K^{*-}$, and $B^-\rightarrow \rho^0\rho^-$.
where SU(3) relations are expected to hold also. In this case, $\omega-\phi$
mixing is ``ideal'' mixing with $\phi$ a pure $s\bar s$ state.

One might think the same analysis can be identically applied to $B^-\rightarrow
\rho^- \bar K^0\;, \rho^0 K^-\;, \omega K^-$ and $B^-\rightarrow \rho^-\pi^0$,
or $B^-\rightarrow
\pi^- \bar K^{*0}\;, \pi^0 K^{*-}\;, \eta K^{*-}$ and $B^-\rightarrow
\pi^-\rho^0$, seperately. In each of the above two cases, there is similar
triangle relations analogous to eq.(8) among the first three decay amplitudes.
However, similar
relations to eq.(15,17) are no longer valid. This is because in this case there
is no bose statistics. There are two ways of writing SU(3) invariant
relations, for example, for $H_{\bar 3}$, we can write
\begin{eqnarray}
H_{eff} = C^V_{\bar 3} B_i V^i_l M^j_k H^k_{\bar 3} + C^M_{\bar 3} B_i M^i_l
V^l_kH^k_{\bar 3}\;,
\end{eqnarray}
where $V^i_j$ is the octet-vector meson. If $C^V = C^M$ one would have the
desired relations. However, there is no reasons for $C^V$ and $C^M$ to be
equal.
Similarly for other SU(3) invariant amplitudes. Because of lack of this
equality, similar relations to eqs. (15,17) do not exist.

In conclusion we have proposed a new method to determine the phase $\gamma$
which is free from contamination by electroweak penguin contributions.
All decays involved
have branching ratios of order $10^{-5}$, and are in principle measurable at
existing or future $B$ facilities.

\begin{figure}[htb]
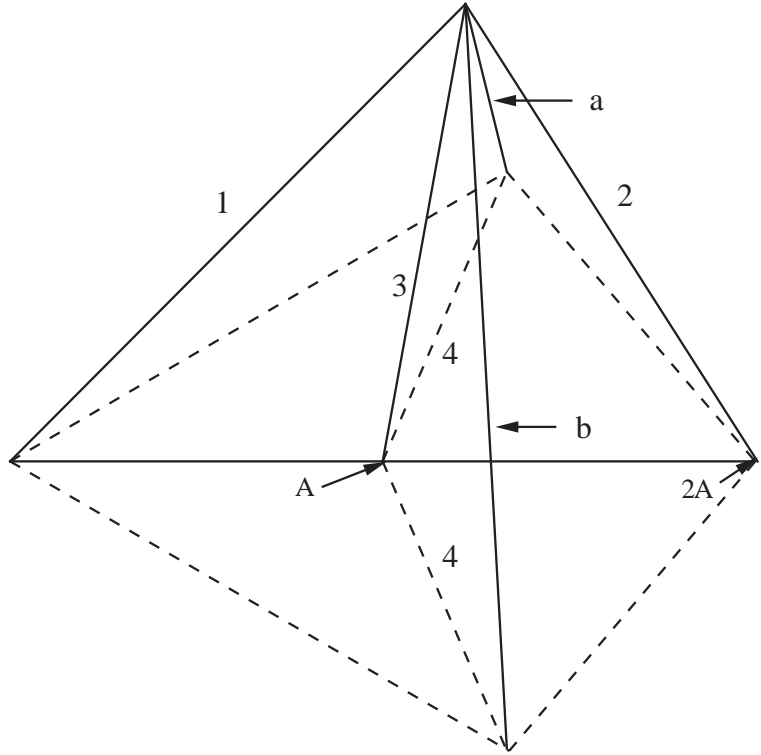

\centerline{ \DESepsf(gamma.epsf width 10 cm) }
\smallskip
\caption {The triangle relations and the two solutions a and b for $B-\bar B$.
Lines A, 1, 2, 3, 4 are the amplitudes
$A(\pi^-\bar K^0)$,
$\sqrt{2} A(\pi^0 K^-)$,
$\sqrt{6} A(\eta_8 K^-)$,
$B$ and $\bar B$,respectively.
The dashed lines are for the corresponding anti-B decay amplitudes.}
\label{gamma}
\end{figure}

\end{document}